\documentclass{PoS}

\newcommand{\be}{\begin{equation}}
\newcommand{\ee}{\end{equation}}
\newcommand{\bea}{\begin{eqnarray*}}
\newcommand{\eea}{\end{eqnarray*}}
\newcommand{\bs}{\begin{structureenv}}
\newcommand{\es}{\end{structureenv}}
\newcommand{\call}{{\cal C}}

\newcommand{\vpot}{{\mathit a}}
\newcommand{\LL}{{\cal L}}

\title{From Loops to Surfaces}

\ShortTitle{From Loops to Surfaces}

\author{\speaker{H. Neuberger}\thanks{HN acknowledges partial support by the DOE under grant number DE-FG02-01ER41165. }\\Rutgers University, Department of Physics and Astronomy, Piscataway, NJ 08855, USA\\
        E-mail: \email{neuberg@physics.rutgers.edu}}

\author{R. Narayanan\thanks{R.N. acknowledge partial support by the NSF under
grant number PHY-0854744. }\\Florida International University, Department of Physics,  Miami, FL 33199, USA\\E-mail: \email{rajamani.narayanan@fiu.edu}}

\abstract{The generating function for all antisymmetric characters of a 
Wilson loop matrix in $SU(N)$ Yang Mills theory is the partition function
of a fermion living on the curve describing the loop. This generalizes to fermion 
subsystems living on higher dimensional submanifolds, for example, surfaces. 
This write-up also contains some 
extra background, in response to some questions raised during the oral presentation.}

\FullConference{The XXVIII International Symposium on Lattice Field Theory, Lattice2010\\
		June 14-19, 2010\\
		Villasimius, Italy}

\begin{document}
\section{Introduction}
In D Euclidean dimensions, at $N=\infty$ $SU(N)$ pure YM simplifies~\cite{thooftN}, but this has not yet been turned 
into a quantitative tool. We are trying to improve on this situation.

Let ${\cal O}$ be an observable characterized by one single scale $l$ 
in 4D. 
For $l\Lambda_N\ll 1$, $\langle{\cal O}\rangle_{N=\infty}$ can be 
computed by summing up contributions from planar Feynman diagrams.
Moreover, there are good reasons to believe that the series in $\frac{1}{\log( l\Lambda_N )}$ converges~\cite{thooftcv}.
$\Lambda_N$ is the analogue of $\Lambda_{\rm QCD}$ for pure $SU(N)$ gauge theory. 
For $l\Lambda_N\gg 1$, for a class of specific ${\cal O}$'s, $\langle{\cal O}\rangle_{N=\infty}$ can be expanded in $\frac{1}{l\Lambda_N}$ using an 
effective theory based on free strings~\cite{aharony}. 

How are the two regimes connected ?

Some ${\cal O}$'s have a narrow crossover as $l$ changes from $l\Lambda_N \lesssim 1$
to $l\Lambda_N\gtrsim 1$ at finite $N$, becoming a ``phase transition'' at 
$N=\infty$. Examples of such ${\cal O}$'s
will be presented later -- in these examples the operators turn out to be non-local~\cite{oura}. 
Such large $N$ phase transitions tend to fall in Random Matrix universality classes. The hope is to exploit this to connect field theoretical
perturbation theory to effective string theory. 

\section{Loops}
\subsection{Wilson loop operator -- ignoring renormalization}

Let $R$ be an irreducible $SU(N)$ representation of dimension $d_R$. We define the nonlocal matrix valued
operator:
\be
{\cal P} e^{i\oint_{\cal C} A_R \cdot dx}\equiv \Omega_R({\cal C})
\ee
${\cal P}$ denotes ordering round the closed
curve ${\cal C}$. The starting/end point of the ordering is suppressed because
we shall build out of $\Omega_R({\cal C})$ other operators which will all be
independent of the choice of this point on ${\cal C}$. 
The Wilson loops are the following averages:
\be
W_R({\cal C})={\rm tr}\langle \Omega_R({\cal C})\rangle/d_R
\ee

Typically, one focuses only on the fundamental representation, which we denote by $R=1$ or $R=f$. 
We depart from this by considering the collection of all fully antisymmetric representations,
$R=k, N-k$; $k=1,...,\left [\frac{N}{2}\right ]$. The generating function for the $W_k({\cal C})$ is $\langle \det (z+\Omega_f ({\cal C} )) \rangle=Q(z,\call)$, a 
palindromic polynomial of rank $N$ in $z$. This permits us to introduce a set of quantities we
loosely refer to as the ``eigenvalues'' of $\Omega_1({\cal C})$, the roots of $Q(z,\call)$. 
As the loop ${\cal C}$ is dilated, its minimal area $A$ grows and one can extract $k$-string tensions 
$\sigma_k$ from $W_k({\cal C})\sim \exp(-\sigma_k A)$. 

The following scenario is plausible: 
A large $N$ ``phase transition'' (meaning: a non-smooth change in the dependence on $z$) separates 
small from large loops and occurs in D=2,3,4. 
As $N\to\infty$ a continuum density 
of ``eigenvalues'', roots of $Q(z,\call)$, develops, supported on $|z|=1$,
gapped at $z=1$ for small loops, and uniform for infinite loops.
Hence, for very large loops the deviation of the density 
from uniformity is controlled by $[N/2]$ exponents
$e^{-\sigma_k A}$, dominated by $e^{-\sigma_f A}$. 
Here we imposed an assumed parity invariance.

Close to critical loop-size, and for $z$ close to $1$, there is a universal description common to all dimensions.
The case $D=2$ is exactly soluble so the universal form is known~\cite{ourb}. 
This phase transition is seen only in $Q(z,\call)$, but not in the individual $W_k$'s and
is therefore different from the familiar Gross-Witten phase transition~\cite{gw}. 
The universality provides an economic parametrization of the 
short-scale to long-scale crossover in $Q(z,\call)$ for $1\ll N <\infty$.

For this to be meaningful we need to renormalize $Q(z,\call)$.
After this is done, we can use lattice gauge theory simulations to validate the scenario
as a property of continuum large $N$ $SU(N)$ gauge theory.

\subsection{Renormalization of Wilson loops}

Our bare observable is $Q(z,\call)=\langle \det (1+z \Omega_f^\dagger ({\cal C}) )\rangle$.  The following representation is an exact identity for an 
$\Omega_f$ given by a fixed $N\times N$ unitary matrix of unit determinant~\cite{ourc}:
\be
\det[1+z\Omega_f^\dagger (\call ) ]=\int [d\bar\psi d\psi ] e^{\int_0^l d\sigma
\bar\psi(\sigma)[\partial_\sigma - \mu - i\vpot(\sigma)]\psi(\sigma)}.
\ee
Here, $z=e^{-\mu l}$, 
$\sigma$ parametrizes $\call$ by $x(\sigma)$ with 
$[\partial_\sigma x_\mu (\sigma )]^2=1$ and $l$ is the length of $\call$.
The Grassmann variables $\bar\psi(\sigma),\psi(\sigma)$ obey anti-periodic boundary
conditions. The gauge potential along the curve is projected from the D--dimensional 
gauge potential by $\vpot (\sigma) = A_\mu (x(\sigma ))
\frac {\partial x_\mu(\sigma)}{d\sigma}.$

We rewrite out bare observable accordingly:
\be
Q(z,\call)=
\langle \int [d\bar\psi d\psi ] e^{\int_0^l d\sigma
\bar\psi(\sigma)[\partial_\sigma - \mu - i\vpot(\sigma)]\psi(\sigma)}\rangle
\ee
Now we employ power counting to determine the needed counterterms (ct-s) for renormalization~\cite{ourd}.
The parameterization choice of the curve makes the curve 
parameter of dimension inverse mass,
$[\sigma]=-1$, which in turn makes the Grassmann fields unitless, $[\bar\psi,\psi]=0$.
Terms containing more than one $\bar\psi\psi$ pair and a single derivative 
can be eliminated by a field redefinition. This leaves the non-redundant ct-s, $[{\bar \psi}\psi]^k,\;k=1,..N$. 
The symmetry 
$\psi \to\bar\psi,~~ \bar\psi \to \psi,~~ A_\mu \to A_\mu^\ast$
can be preserved reducing the 
number of ct-s to $\left [\frac{N}{2}\right ]$.

This leaves exactly the amount of freedom needed to arrange for the 
inequality $\frac{1}{d_R} W_R (\call ) \le 1$ to hold for all antisymmetric $R$, 
after renormalization. 
So long as this is true, some further 
adjustments (not finely tuned) of the finite parts of the coefficients of the ct-s
will render  $Q(z,\call)$ palindromic with all roots on $|z|=1$.
To be sure, in 4D and 3D, the short distance 
singularity that one expects from one gluon exchange even 
after the removal of the perimeter divergence, is eliminated by the
integration round the contour and one can
set the expectation values to their classical values for zero size loops. 
In 4D for example, the one gluon exchange
answer could be regularized to give a nontrivial 
conformally invariant functional of
the loop shape~\cite{stodo}, but we do not do this here. 

The divergences one has to eliminate are linear in 4D and logarithmic in 3D. 
There are no divergences in 2D. All our loops are assumed to be differentiable, so the corner singularities of the 4D case do not occur. 

\subsection{A Dirac operator associated with the large $N$ transition}

The spectrum of $D_1 (\call)\equiv \partial_\sigma - i\vpot(\sigma) $ 
will have a gap for small loops and will be gap-less for large loops.
There is an analogy to spontaneous chiral symmetry breaking and 
its connection
to chiral random matrix theory.
This analogy is exploited in our generalization to surfaces. 

\subsection{More about the divergences}

The $\left [\frac{N}{2}\right ]$ ct-s are necessary to eliminate the
$\left [\frac{N}{2}\right ]$ perimeter divergences associated with the distinct
$N$-ality representations, not counting conjugate ones.
Physically, the ct-s represent the arbitrary amounts of thickening the distinct
$k$-strings need.

On the lattice one ``thickens'' the gauge fields the fermions see.
For example, in APE smearing \cite{ape10} one defines a recursion 
starting from the link matrices $U_\mu(x)$ at step $n=0$ and proceeds from step
$n$ to step $n+1$ by:
\begin{eqnarray}
X^{(n+1)}_\mu (x;f)= (1-|f|) U^{(n)}_\mu (x;f)+\frac{f}{2(d-1)}
\Sigma_{U^{(n)}_\mu (x;f)}\nonumber\\ U^{(n+1)}_\mu (x; f
)=X^{(n+1)}_\mu (x;f) \frac{1}{\sqrt{[X^{(n+1)}_\mu (x;f)]^\dagger
X^{(n+1)}_\mu (x;f)}}
\end{eqnarray}
$\Sigma_{U^{(n)}_\mu (x;f)}$ denotes the ``staple''
associated with the link $U^{(n)}_\mu(x;f)$ in terms of the entire
set of $U^{(n)}_\nu(y;f)$ matrices. Writing $U^{(n)}_\mu
(x;f)=\exp(iA^{(n)}_\mu (x;f))$, and expanding in $A_\mu$ one
finds~\cite{pertsmear11}, in lattice Fourier space:
\be
A^{(n+1)}_\mu (q;f)= \sum_\nu h_{\mu\nu} (q) A^{(n)}_\nu (q;f) \ee
with
\be
h_{\mu\nu} (q)= f(q)(\delta_{\mu\nu} - \frac {{\tilde q}_\mu
{\tilde q}_\nu}{\tilde q^2})+\frac {{\tilde q}_\mu {\tilde
q}_\nu}{\tilde q^2} \ee where $\tilde q_\mu = 2\sin
(\frac{q_\mu}{2})$ and
\be
f(q)=1-\frac{f}{2(d-1)}\tilde q^2
\ee
The iteration is solved by
replacing $f(q)$ by $f^n (q)$, where, for small enough $f$,
\be
f^n(q) \sim e^{-\frac{f n}{2(d-1)} \tilde q^2}
\label{linear}
\ee
One can think about $f^n(q)$ as a gauge 
invariant, Euclidean, form factor in a
particular gauge; this form-factor is 
attached to the gauge field -- fermion
pair vertex. 

One can take a trivial continuum limit~\cite{wilev}
on the step number and $f n$ is then seen to play the role of
the fifth ``time'', $\tau$, in a Langevin equation where the
noise has been set to zero. The appropriate D+1 dimensional gauge
covariant set of partial differential equations is
\be
F_{5\nu} =D_\mu F_{\mu\nu}, \ee
where  we use continuum notation also for the gauge fields and have set $\tau=x_5$.
These equations have only an $O(D)$ invariance, not $O(D+1)$. They
are first order in $x_5$ and second order in $x_\mu ,\mu=1,2,3,4$. 
Gauge fixing to $A_5=0$ and going back to $\tau$, we get
\be
\frac{\partial A_\mu}{\partial\tau} =D_\mu F_{\mu\nu}
\label{langev}
\ee
with a residual 4D gauge invariance -- see \cite{wilev} for references to earlier 
work on similar equations. 

The parameter $\tau$ tells us, in units of length-squared
~(\ref{linear}), the rough amount of regularizing ``fat'' that has been 
added to the loop. All ultraviolet divergences of Wilson 
loop operators are eliminated as soon as $\tau >0$.
These would include corner divergences if there were any; on the lattice
there always will be. 

We take $\tau$ proportional to the physical perimeter of the loop squared 
for small loops, and approaching a constant length squared (relatively small 
relative to $\Lambda_N^{-2}$) for large loops.
This is done in order to remain sensitive to the short distance structure of the theory for
small loops while allowing large loops to have only a modest amount of thickness. 
The confinement of the fat loops will still be governed by exponential terms with the same $\sigma_k$ one would have extracted from bare Wilson loops, which were not fattened. 

\section{Surfaces}

The need to smear the loops and to adjust a number of order $N$ of finite 
parts renders the renormalized $Q(z,{\cal C})$ observable somewhat inelegant.
This leads us to the idea of looking at less singular objects than lines -- less
singular means that more displacement directions are allowed. 
The simplest step in this direction is to replace the curve by a 2-dimensional 
surface $\Sigma$, described by $x_\mu(\sigma)~(\sigma_\alpha,\alpha=1,2)$. We put massive 
Dirac fermions on $\Sigma$, which is characterized by a single scale $l$.
At zero mass, one has chiral symmetry 
and the fermionic determinant is the exponent of the Polyakov--Wiegmann action
~\cite{polywieg}.  
The gauge connection on $\Sigma$ is $\vpot_\alpha=A_\mu (x(\sigma))\frac{\partial x_\mu}{\partial \sigma_\alpha}$.
The massless Dirac operator is 
$D_2 (\Sigma )=\gamma_\alpha [\partial_{\sigma_\alpha} -  
i\vpot_\alpha(\sigma)]$ and our new observable becomes 
\be
Q(\mu,\Sigma)=\langle
\int [d\bar\psi d\psi ]  
e^{\int_\Sigma d^2\sigma
\bar\psi(\sigma)[D_2(\Sigma)-\mu ] \psi(\sigma)}\rangle\ee

The possible counterterms are built out of currents on the surface: $J^j_\alpha (\sigma ) = \bar\psi (\sigma) \gamma_\alpha T^j \psi (\sigma)$, $J_\alpha = \bar\psi (\sigma) \gamma_\alpha \psi (\sigma)$.
So, only two ct-s, $\LL_1 = J^j_\alpha J^j_\alpha,~~~\LL_2= J_\alpha J_\alpha$ 
are now available. 
In 4D one has logarithmic divergences, and 
${\cal L}_1$ will be generated, but in 3D no ct-s are generated because the
theory is superrenormalizable. 

At $N=\infty$ a S$\chi$SB transition is possible, 
depending on $l$; for a finite surface, this
is a large $N$ phase transition, analogue to the one occurring in the smeared Wilson loops. For finite $N$ this transition disappears. 

The lattice version of $Q(\mu,\Sigma )$ is straightforward to construct: 
we used the overlap operator for $D_2$~\cite{overlap} to ensure a precise
definition of the location of spontaneous chiral symmetry breaking on the lattice.

\section{Results in 3D}

As said, no extra 
renormalization necessary in 3D. We set $\mu=0$ and $N=\infty$.
For an infinite plane we find $\langle \bar\psi\psi\rangle_{N=\infty} =0.29(1) \sqrt{\sigma_f }$, 
where $\sigma_f$ is the $k=1$ string tension in the pure YM theory. 
For a cylinder with square base of side $s$ we establish that spontaneous chiral symmetry breaking occurs when $s > s_c$,  while for $s<s_c$ chiral symmetry is preserved. 
For $s > s_c$ with $(s-s_c)/s_c\ll 1$ we have 
$\langle\bar\psi\psi\rangle_{N=\infty}(s) \propto (s-s_c)^{1/2}$. 
The critical size is found to be $s_c= 1.4(1)/\sqrt{\sigma_f}$.

\section{Outlook}

In 3D the definition of the curve related
observable that exhibited the large $N$ phase transition required extra regularization which we implemented as smearing. Moving from curves to surfaces
eliminated this smearing. In 4D there are
further obstacles which we have to learn how to overcome. 
However, the number of ct-s is independent of $N$ now. 

In 3D, we described above 
two unambiguously defined $N=\infty$ scales, $\langle\bar\psi\psi\rangle$ and $s_c$.
We are looking for analytical ways to 
estimate these scales.


\begin{thebibliography}{99}

%\cite{'tHooft:1973jz}
\bibitem{thooftN}
  G.~'t Hooft,
  %``A PLANAR DIAGRAM THEORY FOR STRONG INTERACTIONS,''
  Nucl.\ Phys.\  B {\bf 72}, 461 (1974).
  %%CITATION = NUPHA,B72,461;%%


%\cite{'tHooft:1982cx}
\bibitem{thooftcv}
  G.~'t Hooft,
  %``Rigorous Construction Of Planar Diagram Field Theories In Four-Dimensional
  %Euclidean Space,''
  Commun.\ Math.\ Phys.\  {\bf 88}, 1 (1983).
  %%CITATION = CMPHA,88,1;%%

%\cite{Aharony:2009gg}
\bibitem{aharony}
  O.~Aharony and E.~Karzbrun,
  %``On the effective action of confining strings,''
  JHEP {\bf 0906}, 012 (2009)
  [arXiv:0903.1927 [hep-th]],
  %%CITATION = JHEPA,0906,012;%%


%\cite{Aharony:2010db}
%\bibitem{Aharony:2010db}
  O.~Aharony and N.~Klinghoffer,
  %``Corrections to Nambu-Goto energy levels from the effective string action,''
  arXiv:1008.2648 [hep-th],
  %%CITATION = ARXIV:1008.2648;%%

%\cite{Aharony:2010cx}
%\bibitem{Aharony:2010cx}
  O.~Aharony and M.~Field,
  %``On the effective theory of long open strings,''
  arXiv:1008.2636 [hep-th].
  %%CITATION = ARXIV:1008.2636;%%

%\cite{Aharony:2010af}







\bibitem{oura} R.~Narayanan and H.~Neuberger,
%  ``Large N reduction in continuum,''
  Phys.\ Rev.\ Lett.\  {\bf 91}, 081601 (2003).
%  [arXiv:hep-lat/0303023]
 %\cite{Narayanan:2008he}
%\bibitem{Narayanan:2008he}
  R.~Narayanan, H.~Neuberger and E.~Vicari,
  %``A large N phase transition in the continuum two dimensional SU(N) X SU(N)
  %principal chiral model,''
  JHEP {\bf 0804}, 094 (2008)
  [arXiv:0803.3833 [hep-th]].
  %%CITATION = JHEPA,0804,094;%%
%\cite{Narayanan:2007ug}
%\bibitem{Narayanan:2007ug}
  R.~Narayanan, H.~Neuberger and F.~Reynoso,
  %``Phases of three dimensional large N QCD on a continuum torus,''
  Phys.\ Lett.\  B {\bf 651}, 246 (2007)
  [arXiv:0704.2591 [hep-lat]].
  %%CITATION = PHLTA,B651,246;%%



 %\cite{Lohmayer:2009aw}
\bibitem{ourb}
  R.~Lohmayer, H.~Neuberger and T.~Wettig,
  %``Eigenvalue density of Wilson loops in 2D SU(N) YM,''
  JHEP {\bf 0905}, 107 (2009)
  [arXiv:0904.4116 [hep-lat]].
  %%CITATION = JHEPA,0905,107;%%
%\cite{Neuberger:2008mk}
%\bibitem{Neuberger:2008mk}
  H.~Neuberger,
  %``Burgers' equation in 2D SU(N) YM,''
  Phys.\ Lett.\  B {\bf 666}, 106 (2008)
  [arXiv:0806.0149 [hep-th]].
  %%CITATION = PHLTA,B666,106;%%
%\cite{Neuberger:2008ti}
%\bibitem{Neuberger:2008ti}
  H.~Neuberger,
  %``Complex Burgers' equation in 2D SU(N) YM,''
  Phys.\ Lett.\  B {\bf 670}, 235 (2008)
  [arXiv:0809.1238 [hep-th]].
  %%CITATION = PHLTA,B670,235;%%
%\cite{Narayanan:2007dv}
%\bibitem{Narayanan:2007dv}
  R.~Narayanan and H.~Neuberger,
  %``Universality of large N phase transitions in Wilson loop operators in two
  %and three dimensions,''
  JHEP {\bf 0712}, 066 (2007)
  [arXiv:0711.4551 [hep-th]].
  %%CITATION = JHEPA,0712,066;%%



%\cite{Gross:1980he}
\bibitem{gw}
  D.~J.~Gross and E.~Witten,
  %``Possible Third Order Phase Transition In The Large N Lattice Gauge
  %Theory,''
  Phys.\ Rev.\  D {\bf 21}, 446 (1980).
  %%CITATION = PHRVA,D21,446;%%


%\cite{Narayanan:2010zg}
\bibitem{ourc}
  R.~Narayanan and H.~Neuberger,
  %``Two dimensional fermions in three dimensional YM,''
  JHEP {\bf 1006}, 014 (2010)
  [arXiv:1005.0576 [hep-lat]].
  %%CITATION = JHEPA,1006,014;%%


\bibitem{ourd}
%\cite{Narayanan:2009ag}
%\bibitem{Narayanan:2009ag}
  R.~Narayanan and H.~Neuberger,
  %``Two dimensional fermions in four dimensional YM,''
  JHEP {\bf 0911}, 018 (2009)
  [arXiv:0909.4066 [hep-lat]].
  %%CITATION = JHEPA,0911,018;%%



\bibitem{stodo} 
  L.~Stodolsky,
  %``Classical radiation of a finite number of photons,''
  Acta Phys.\ Polon.\  B {\bf 33}, 2659 (2002)
  [arXiv:hep-th/0205313].
  %%CITATION = APPOA,B33,2659;%%



\bibitem{ape10}
T. DeGrand, Phys. Rev. D63 (2001) 034503;
M. Albanese et. al. [APE Collaboration], Phys.\ Lett.\ B{\bf 192}, (1987) 163;
M. Falcioni, M.L. Paciello, G. Parisi and B. Taglienti, Nucl. Phys. Nucl.\ Phys.\ {\bf
B251}, 624 (1985).
\bibitem{pertsmear11}
C.~W.~Bernard and T.~DeGrand,
  %``Perturbation theory for fat-link fermion actions,''
  Nucl.\ Phys.\ Proc.\ Suppl.\  {\bf 83}, 845 (2000).



\bibitem{wilev} 
R.~Narayanan and H.~Neuberger,
%  ``Infinite N phase transitions in continuum Wilson loop operators,''
  JHEP {\bf 0603}, 064 (2006).
%  [arXiv:hep-th/0601210]
 
\bibitem{polywieg}

%\cite{Polyakov:1984et}
%\bibitem{Polyakov:1984et}
  A.~M.~Polyakov and P.~B.~Wiegmann,
  %``Goldstone Fields In Two-Dimensions With Multivalued Actions,''
  Phys.\ Lett.\  B {\bf 141}, 223 (1984),
  %%CITATION = PHLTA,B141,223;%%

%\cite{Polyakov:1983tt}
%\bibitem{Polyakov:1983tt}
  A.~M.~Polyakov and P.~B.~Wiegmann,
  %``Theory of nonabelian Goldstone bosons in two dimensions,''
  Phys.\ Lett.\  B {\bf 131}, 121 (1983).
  %%CITATION = PHLTA,B131,121;%%

\bibitem{overlap} 
H.~Neuberger,
%  ``Exactly massless quarks on the lattice,''
  Phys.\ Lett.\  B {\bf 417}, 141 (1998); 
% H.~Neuberger,
%  ``More about exactly massless quarks on the lattice,''
  Phys.\ Lett.\  B {\bf 427}, 353 (1998). 
%  [arXiv:hep-lat/9801031]



\end{thebibliography}
\end{document}